\pdfoutput=1
\documentclass[preprint, 5p, times]{elsarticle}
\usepackage{graphicx}
\usepackage{amsmath, amssymb}

\usepackage[utf8]{inputenc}
\usepackage[T1]{fontenc}

\usepackage[protrusion=true,expansion=true]{microtype}

\makeatletter
\def\ps@pprintTitle{%
 \let\@oddhead\@empty
 \let\@evenhead\@empty
 \def\@oddfoot{\footnotesize\itshape 
       Revised manuscript subsequently published as \href{http://dx.doi.org/10.1016/j.ijms.2014.01.028}{DOI:10.1016/j.ijms.2014.01.028} \hfill \today}%
 \let\@evenfoot\@oddfoot}
\makeatother

\usepackage[pdftex, bookmarks = true]{hyperref}
\hypersetup{
pdftitle = {Classical calculation of relativistic frequency-shifts in an ideal Penning trap}, 
pdfauthor = {Jochen Ketter et al.}, 
pdfsubject = {Accepted manuscript}, 
pdfkeywords = {Penning trap; mass spectrometry; perturbation theory; special relativity}, 
pdfstartview = {FitH},
}

\journal{International Journal of Mass Spectrometry}
\biboptions{sort&compress}

\newcommand*{\myd}{\mathrm{d}} 

\newcommand*{\myGSAbstand}{\ } 

\newcommand*{\myAmpR}[1]{\hat{\rho}_{#1}} 

\newcommand*{\myAmpZ}{\hat{z}} 

\newcommand*{\myvec}[1]{\vec{#1}} 

\newcommand*{\mybeat}{\mathrm{b}} 

\newcommand*{\myeuv}[1]{\vec{e}_{#1}} 

\newcommand*{\myomc}{\omega_\mathrm{c}} 

\newcommand*{\myomz}{\omega_z} 

\newcommand*{\myomzw}{\tilde{\omega}_{z}} 

\newcommand*{\myomm}{\omega_{-}} 

\newcommand*{\myommw}{\tilde{\omega}_{-}} 

\newcommand*{\myomp}{\omega_+} 

\newcommand*{\myompw}{\tilde{\omega}_{+}} 

\newcommand*{\myompm}{\omega_{\pm}} 

\newcommand*{\myommp}{\omega_{\mp}} 

\newcommand*{\myomw}{\tilde{\omega}} 

\newcommand*{\myompmw}{\tilde{\omega}_{\pm}} 

\newcommand*{\myommpw}{\tilde{\omega}_{\mp}} 

\newcommand*{\myombw}{\tilde{\omega}_{\mybeat}} 

\newcommand*{\mynabla}{\vec{\nabla}} 

\newcommand*{\myfComp}[2]{\left\langle {#1} \right\rangle_{#2}} 

\newcommand*{\myfZero}{0} 

\newcommand*{\myPhiTw}[1]{\tilde{\omega}_{#1}t+\tilde{\varphi}_{#1}} 

\newcommand*{\myPhiTotw}[1]{\tilde{\chi}_{#1}} 

\newcommand*{\myZOT}[1]{\tilde{#1}} 

\newcommand*{\myels}{\varepsilon} 

\newcommand*{\mymfls}{\beta} 

\newcommand*{\myvel}{\varv} 

\newcommand*{\myEn}[1]{\mathcal{E}_{#1}} 
\newcommand*{\mySec}[1]{Section~\ref{#1}}
\newcommand*{\myEqua}[1]{Equation~\eqref{#1}}
\newcommand*{\myEquas}[1]{Equations~\eqref{#1}}
\newcommand*{\myTab}[1]{Table~\ref{#1}}

\begin{document}
\begin{frontmatter}

\title{Classical calculation of relativistic frequency-shifts in an ideal Penning trap}

\author{Jochen Ketter\corref{cor1}}
\ead{jochen.ketter@mpi-hd.mpg.de}
\cortext[cor1]{Corresponding author}
\author{Tommi Eronen\corref{cor2}}
\author{Martin Höcker}
\author{Marc Schuh}
\author{Sebastian Streubel}
\author{Klaus Blaum}

\address{Max-Planck-Institut für Kernphysik, Saupfercheckweg 1, 69117 Heidelberg, Germany}

\begin{abstract}
The ideal Penning trap consists of a uniform magnetic field and an electrostatic quadrupole potential. 
In the classical low-energy limit, the three characteristic eigenfrequencies of a charged particle trapped in this configuration do not depend on the amplitudes of the  three eigenmotions. 
No matter how accurate the experimental realization of the ideal Penning trap, its harmonicity is ultimately compromised by special relativity. 
Using a classical formalism of first-order perturbation theory, we calculate the relativistic frequency-shifts associated with the motional degrees of freedom for a spinless particle stored in an ideal Penning trap, and we compare the results with the simple but surprisingly accurate model of relativistic mass-increase.
\end{abstract}



\end{frontmatter}

\section{Introduction} 
Despite its versatility~\cite{blaum2010} and the eigenmotion called the modified cyclotron-mode, the Penning trap is not perceived as an accelerator---a device typically viewed as capable of producing highly-energetic particles for which relativistic mass-increase plays an important role. 
Given the small scale of the Penning trap ranging from millimeters to a few centimeters, the charged particle stored in it may appear to move in the purely classical domain, well outside the realm of special relativity. 
In the classical limit, the three eigenfrequencies---all of which depend on the mass of the stored particle to a varying extent---are independent of the motional amplitudes. 
However, apart from possibly being too small to be detected, there is no threshold for the onset of relativistic effects and hence even the ideal Penning trap is inherently anharmonic. 
It is because of the outstanding precision of up to $10^{-10}$ for single frequency measurements that a relativistic shift was crucial to the determination of the antiproton mass~\cite{gabrielse1995}. 
Similarly, relativistic shifts may be dominant sources of uncertainty in measurements on  light or highly-charged ions~\cite{bergstroem2002,brodeur2009,sturm2011}. 
Conversely, these shifts are particularly interesting for measuring motional amplitudes~\cite{redshaw2006,myers2013} because, unlike the anharmonic shifts caused by other imperfections, they do not depend on specific parameters of the trap apart from the readily measured frequencies.

Probably because of early work on electrons and the interest in their magnetic moment~\cite{vanDyck1986}, the theoretical treatment of relativistic frequency-shifts used quantum-mechanical operator formalisms~\cite{graeff1969,ringhofer1974,brown1986,cptI2005}. 
When relativistic equations of motion were considered~\cite{kaplan1982,gabrielse1985,guan1993}, the focus was more on excitations of the modified cyclotron-mode than on static frequency-shifts. 

A classical treatment with relativistic additions does not have to be conceptually inferior to a fully relativistic or quantum-mechanical approach, in particular if quantization remains unobservable and an exact solution is impossible in either case. 
In fact, reproducing the classical limit is in general a benchmark for a relativistic quantum theory.
Consequently, knowing the prediction of a non-quantized treatment is worthwhile. 

In this paper, we show that the relativistic frequency-shifts caused by the motional degrees of freedom of a charged particle stored in an ideal Penning trap are also reproduced in a classical framework of perturbation theory. 
With classical framework we refer to the use of equations of motion in contrast to operators and eigenstates. 
In \mySec{sec:TaM}, we approximate the relativistic equation of motion such that classical perturbation theory can be applied with the classical limit of the ideal Penning trap as the starting point. 
We also outline our particular implementation of first-order perturbation theory. 
The actual relativistic frequency-shifts are calculated in \mySec{sec:CalcFreqShifts}. 
In \mySec{sec:EffMass}, the result is then compared with a simple model of relativistic mass-increase.

\section{Theory and method} \label{sec:TaM}
The theoretical framework of perturbation theory is essentially the same as the one we used to calculate the first-order frequency-shifts caused by static cylindrically-symmetric electric and magnetic imperfections of a Penning trap~\cite{ketter2013ijms}. 
This time, we have to learn how to incorporate relativistic effects as a perturbation in the classical equations of motion. 
To this end, we take a more general look at the relativistic equations of motion in search of a suitable perturbation parameter, before plugging in the specific electric and magnetic field of the ideal Penning trap. 

\subsection{Relativistic equation of motion} \label{subsec:EQM}

Consider a static electric field~$\myvec{E}$ and a static magnetic field~$\myvec{B}$ in the laboratory frame. We will use this frame exclusively throughout the paper, never looking at the particle's rest frame or its proper time. 
Accordingly, all time-derivatives shown are with respect to time in the laboratory frame. 
Moreover, the limitation to static fields spares us from the complications of retardation. 
We will also ignore radiation damping because the emission of synchrotron radiation is insignificant for particles heavier than electrons~\cite{brown1986}.
For a pointlike spinless particle of charge~$q$ and rest mass~$m$, the relativistic equation of motion is then given by
\begin{gather}
\dot{\myvec{p}} 
= \frac{\myd}{\myd t}\myvec{p} 
= \frac{\myd}{\myd t} (\gamma m \myvec{\myvel}) 
= q(\myvec{E} + \myvec{\myvel}\times \myvec{B}) 
\label{eq:RelativisticEQM} \myGSAbstand ,
\end{gather}
where $\myvec{p}$ is the particle's momentum and $\myvec{\myvel}$ its velocity. 
We will use $p = \lvert\myvec{p}\rvert$ and $\myvel = \vert\myvec{\myvel}\rvert$ as an abbreviation for the length of the corresponding vectors. 
Thus far, the Lorentz factor~$\gamma = 1/\sqrt{1-\myvel^2/c^2}$ with the speed of light~$c$ is the only difference from the classical Newtonian equation of motion. 
However, in addition to the familiar acceleration~$\dot{\myvec{\myvel}}$, taking the time-derivative of the relativistic momentum~$\myvec{p} = \gamma m \myvec{\myvel}$ results in a time-derivative of the Lorentz factor, which is expressed more conveniently via the particle's total energy $\mathcal{E} = \gamma mc^2$ and the relativistic energy--momentum relation as
\begin{align}
\dot{\gamma} %
= \frac{1}{mc^2}\frac{\myd}{\myd t}\mathcal{E} %
& = \frac{1}{mc^2} \frac{\myd}{\myd t} %
\left[\sqrt{(mc^2)^2+(pc)^2}\right] %
= \frac{\dot{\myvec{p}}\cdot\myvec{p}}{\gamma m^2 c^2} %
\myGSAbstand.
\end{align}
By plugging in the right-hand side of \myEqua{eq:RelativisticEQM} for~$\dot{\myvec{p}}$ and by recalling that the momentum~$\myvec{p}$ is always perpendicular to the force~$q\myvec{\myvel}\times \myvec{B}$ associated with the magnetic field, the relativistic equation of motion is rewritten as
\begin{gather}
\dot{\myvec{\myvel}} = \frac{q}{\gamma m}
\left(\myvec{E}+\myvec{\myvel}\times \myvec{B}\right)
{} - \frac{q}{\gamma mc^2} \myvec{\myvel} \,
\left(\myvec{E}\cdot \myvec{\myvel}\right) 
\label{eq:RelEQM} \myGSAbstand.
\end{gather}
Apart from the Lorentz factor~$\gamma$, which might be understood as relativistic mass-increase by redefining the mass as $m \to \gamma m$, there is an additional term that is not present in the classical Newtonian equations of motion. 
However, these are recovered in the classical limit of $c\to \infty$, and consequently $\gamma \to 1$. 

The ideal Penning trap consists of a homogeneous magnetic field~$\myvec{B}_0 = B_0\myeuv{z}$ that is perfectly aligned along the $z$-axis and an electrostatic field 
\begin{gather}
\myvec{E}_{2} 
= -\mynabla\Phi_{2} 
= \frac{V_0}{2d^2}
\begin{pmatrix}x \\ y \\ -2z\end{pmatrix}
\label{eq:E2}
\end{gather}
that is derived from the quadrupole potential
\begin{gather}
\Phi_{2} = \frac{V_0}{2d^2}\left(z^2-\frac{x^2+y^2}{2}\right)
\myGSAbstand .
\end{gather}
The voltage~$V_0$ and the characteristic dimension~$d$ determine the strength of the electric field gradient. 

Whereas we will present an analytic solution for the classical equations of motion in the ideal Penning trap shortly, no such general solution is possible for the fully relativistic case because of the coupling introduced by the Lorentz factor~$\gamma$. 
The situation is identical to the quantum-mechanical case: 
the Schrödinger Hamiltonian is treated analytically in terms of three uncoupled harmonic oscillators~\cite{graeff1967,sokolov1967}, but no exact solution for the relativistic wave equations of a charged particle in a Penning trap is known. 
Either way, approximations have to be made when relativistic effects are taken into account. 
Since the motion of a charged particle stored in a Penning trap is typically only barely relativistic, a perturbative treatment of the lowest-order relativistic corrections suffices. 

With this simplification in mind, we adapt the relativistic equation of motion~\eqref{eq:RelEQM} accordingly, by expanding the inverse of the Lorentz factor~$\gamma$ as
\begin{gather}
\frac{1}{\gamma} 
= \sqrt{1-\myvel^2/c^2} \approx  1 - \frac{\myvel^2}{2c^2} - \cdots 
\end{gather}
for small velocities~$\myvel \ll c$, 
thereby effectively assigning the role of a perturbation parameter to $c^{-2}$. 
By ignoring all the terms of higher order than $c^{-2}$, such as the next order $c^{-4}$, the equation of motion reads
\begin{gather}
\dot{\myvec{\myvel}}
 = \frac{q}{m} 
\left(1-\frac{\myvel^2}{2c^2}\right) 
\left(\myvec{E}+\myvec{\myvel}\times \myvec{B}\right)
{} - \frac{q}{mc^2} \myvec{\myvel}\,
\left(\myvec{E}\cdot \myvec{\myvel}\right) 
\label{eq:RedRelEQM} \myGSAbstand.
\end{gather}
Note that---as an intrinsically relativistic correction---the last term in \myEqua{eq:RelEQM} already came with a factor $c^{-2}$. 
Therefore, already the lowest-order relativistic correction in the Lorentz factor~$\gamma$ results in a term of order~$c^{-4}$, which we have neglected here.

By inserting the electric field~$\myvec{E}_2$ given in \myEqua{eq:E2} and the uniform magnetic field~$\myvec{B}_0 = B_0\myeuv{z}$ of the ideal Penning trap into \myEqua{eq:RedRelEQM}, the approximate equations of motion become
\begin{gather}
\begin{split}
\begin{pmatrix}\ddot{x}\\ \ddot{y}\\ \ddot{z}\end{pmatrix} 
&= 
 \left[1-\frac{\myvel^2}{2c^2}\right] 
\frac{\myomz^2}{2}
\begin{pmatrix}x\\ y \\ -2z \end{pmatrix}
+ \left[1-\frac{\myvel^2}{2c^2}\right] \myomc 
\begin{pmatrix} \dot{y}\\ -\dot{x}\\ 0\end{pmatrix} \\
& \quad
- \frac{1}{c^2} \frac{\myomz^2}{2}
\begin{pmatrix}\dot{x}\\ \dot{y}\\ \dot{z} \end{pmatrix}
\left[\dot{x}x+\dot{y}y-2\dot{z}z\right] 
\label{eq:EffRelEQM} 
\end{split}
\end{gather}
with the velocity squared 
\begin{gather}
\myvel^2 = \dot{x}^2+\dot{y}^2+\dot{z}^2 
\label{eq:v2} 
\end{gather}
given by the quadratic sum of the individual components of the velocity vector~$\myvec{\myvel}$. 
As abbreviations related to the classical case, we have introduced the free-space cyclotron-frequency
\begin{gather}
\myomc 
= \frac{qB_0}{m} 
\label{eq:omc}
\end{gather}
with which the particle would orbit around the magnetic field-lines if there were no electric field, and the axial frequency
\begin{gather}
\myomz = \sqrt{\frac{qV_0}{md^2}} 
\label{eq:omz} 
\end{gather}
with which the particle oscillates along the magnetic field-lines. 
Clearly, trapping requires $qV_0 > 0$.

In the classical limit of $c\to \infty$, the radial motion of the particle consists of two circular modes with frequencies 
\begin{gather}
\myompm = 
\frac{1}{2} 
\left(\myomc\pm \frac{\myomc}{\lvert\myomc\rvert}\sqrt{\myomc^2-2\myomz^2}\right) 
\label{eq:ompm0rad}\myGSAbstand , 
\end{gather}
where $\myomp$ is called the reduced or modified cyclotron-frequency, and $\myomm$ represents the magnetron frequency. 
Trapping requires $\myomc^2 > 2\myomz^2$. 
The frequencies in the ideal classical trap satisfy the relations
\begin{align}
2\myomp\myomm &= \myomz^2 \label{eq:radProd}\myGSAbstand ,\\
\myomp + \myomm &= \myomc \label{eq:sidebandSum}\myGSAbstand ,
\end{align}
which we will use later on.

\subsection{Perturbation theory} \label{subsec:PertTheory}

The steps for applying a first-order perturbative treatment to the relativistic case are essentially the same as outlined in detail for cylindrically-symmetric imperfections~\cite{ketter2013ijms}. 
Therefore, we repeat only the most important identities here. 

The zeroth-order solutions
\begin{align}
\myZOT{x}(t) &= \myAmpR{+} \cos(\myPhiTw{+}) + \myAmpR{-} \cos(\myPhiTw{-}) 
\label{eq:x0}\myGSAbstand, \\
\myZOT{y}(t) &= -\myAmpR{+} \sin(\myPhiTw{+}) - \myAmpR{-} \sin(\myPhiTw{-})
\label{eq:y0}\myGSAbstand,\\
\myZOT{z}(t) &= \myAmpZ\cos(\myPhiTw{z})  
\label{eq:z0}
\end{align}
for the trajectory of the ideal trap with no relativistic effects provide our starting point. 
In the purely classical case, the frequencies~$\myomw_{i}$ are identical to the unperturbed ones defined in \myEquas{eq:omz} and \eqref{eq:ompm0rad}. 
The amplitudes of the radial modes and the axial mode are then given by $\myAmpR{\pm}$ and $\myAmpZ$, respectively. 
The phases~$\tilde{\varphi}_{i}$ are determined by the initial conditions. 
When the relativistic perturbations are considered, the eigenfrequencies shift, which is included by the use of $\myomw_{i}$ instead of $\omega_{i}$. 
Throughout this paper, we define the frequency-shift 
\begin{gather}
\Delta \omega_{i} = \myomw_{i} - \omega_{i}
\end{gather}
as the difference between the relativistic and the nonrelativistic frequency, and we will not always stress explicitly that our result is good to first order only. 

We will insert the zeroth-order solutions from \myEquas{eq:x0}--\eqref{eq:z0} into the first-order equations of motion~\eqref{eq:EffRelEQM}, looking for terms proportional to the zeroth-order solutions. 
These are the terms we will refer to as naturally resonant because their contribution is always in phase with the zeroth-order motion, and they play a crucial role in determining the first-order frequency-shifts. 

However, not all the resulting terms will be proportional to the zeroth-order solutions. 
This is due to the fact that the relativistic additions to the classical equations of motion are nonlinear. 
As the multiplication of oscillatory terms leads to frequency mixing, additional contributions at various sum and difference frequencies of the eigenmodes arise when inserting the zeroth-order solutions. 
Of course, both the sum and the difference frequency may differ from the frequency of an eigenmode, in which case we will refer to both oscillatory terms as nonresonant. 
Even if, say, the sum frequency is equal to the frequency of an eigenmode and thus contributes resonantly, the difference frequency is likely to be nonresonant and vice versa, unless for very specific commensurability conditions of the eigenfrequencies. 
In contrast, we shall see that the naturally-resonant terms appear without any such assumption about the relation between the three eigenfrequencies. 

Whereas the naturally-resonant terms will be absorbed as a first-order frequency-shift, the remaining nonresonant terms serve as a reminder that the zeroth-order ansatz for the trajectory, which is heavily inspired by the exact solution in the classical case, is no longer the complete solution in the relativistic case. 
The discrepancy occurs at first order in the trajectory. 
Fortunately, the first-order frequency-shift is determined entirely by the zeroth-order solution because the nonresonant terms are out of sync with this dominant component of the motion. 
Viewing a resulting nonresonant term as an additional drive on top of otherwise almost harmonic forces, we expect a small oscillatory response by the particle at the driving frequency. 
Since the nonresonant drive is essentially generated by the particle's zeroth-order motion, the additional oscillation may be regarded as a motional sideband. 
To first order in the frequency-shift, the influence of these sidebands on the particle's motional amplitudes at the fundamental eigenfrequencies is safely neglected, and we will keep interpreting $\myAmpR{\pm}$ and $\myAmpZ$ as the amplitudes of the respective eigenmotions. 

Concerning the resonant terms, on which the paper will focus from now on for their exclusive connection with the first-order frequency-shift, our approach is to collect them in effective equations of motion for the zeroth-order solution. 
Apart from a small rescaling of the characteristic frequencies, the effective equations closely resemble the classical case. 
The relativistic frequency~$\myomw_{i}$ in the zeroth-order solution then remains as the parameter that has to be adjusted accordingly in order to fulfill the effective equations.

For the axial mode, the effective equation of motion takes the form
\begin{gather}
\ddot{\myZOT{z}}(t) + \myomz^2(1+\myels_{z}) \,\myZOT{z}(t) = 0 
\label{eq:AxialEQM-Eff} \myGSAbstand, 
\end{gather}
where the resonant relativistic contributions are described by the parameter~$\myels_{z}$.
Reading off the perturbed axial frequency as $\myomzw = \myomz\sqrt{1+\myels_{z}}$, the frequency-shift is then given by
\begin{gather}
\frac{\Delta \myomz}{\myomz} = \frac{\myels_z}{2} 
\label{eq:DeltaOmz} 
\end{gather}
to first order for $\lvert\myels_{z}\rvert \ll 1 $.

The effective equations of motion for the radial modes take the form
\begin{gather}
\begin{pmatrix}\ddot{\myZOT{x}}_\pm\\\ddot{\myZOT{y}}_\pm\end{pmatrix} 
= \myomc(1+\mymfls_{\pm}) 
\begin{pmatrix}\dot{\myZOT{y}}_\pm\\-\dot{\myZOT{x}}_\pm\end{pmatrix}
 + \frac{\myomz^2(1+\myels_\pm)}{2}
\begin{pmatrix}\myZOT{x}_\pm\\\myZOT{y}_\pm\end{pmatrix} 
\label{eq:radEQM-Eff} \myGSAbstand .
\end{gather}
The two terms~$\mymfls_{\pm}$ and $\myels_{\pm}$ comprise the relativistic contributions that cause a first-order frequency-shift. 
For $\mymfls_{\pm} = \myels_{\pm} = 0$, the radial equations of motion of the ideal Penning trap in the classical limit are recovered. 
The perturbed frequencies are expressed as
\begin{align}
 \myompmw &= \myompm
+ \frac{\partial \myompm}{\partial \myomc} \myomc\mymfls_{\pm}
 + \frac{\partial \myompm}{\partial \myomz^2} \myomz^2\myels_\pm + \cdots 
\label{eq:RadFoFsTaylor}\\
&= \myompm
\pm \frac{\myompm \myomc}{\myomp - \myomm}\mymfls_{\pm}
 \mp \frac{\myomp \myomm}{\myomp - \myomm}\myels_\pm + \cdots 
\label{eq:RadFoFs} \myGSAbstand.
\end{align}
by a Taylor expansion of \myEqua{eq:ompm0rad} around the operating point $\mymfls_{\pm} = \myels_{\pm}= 0$ of the ideal classical Penning trap.

\section{Calculation of frequency-shifts} \label{sec:CalcFreqShifts}

Before calculating the actual frequency-shifts, we introduce a piece of notation and we derive some frequently-used identities. 
From the zeroth-order trajectories in \myEquas{eq:x0}--\eqref{eq:z0}, the corresponding zeroth-order velocities follow as
\begin{align}
\dot{\myZOT{x}}(t) &= 
 -\myompw\myAmpR{+} \sin(\myPhiTw{+})
 - \myommw\myAmpR{-} \sin(\myPhiTw{-}) 
\label{eq:xv0}\myGSAbstand, \\
\dot{\myZOT{y}}(t) &= 
 -\myompw\myAmpR{+} \cos(\myPhiTw{+})
 - \myommw\myAmpR{-} \cos(\myPhiTw{-})  
\label{eq:yv0}\myGSAbstand,\\
\dot{\myZOT{z}}(t) &= 
 -\myomzw\myAmpZ\sin(\myPhiTw{z})  
\label{eq:zv0} \myGSAbstand .
\end{align}
We will use 
\begin{gather}
\myPhiTotw{i} = \myPhiTw{i}
\end{gather}
as a short-hand notation for the total phase without always stressing its time-dependence, just like we will often not show the time-dependence of the zeroth-order solutions for the sake of space.

Because of the two relativistic corrections associated with the velocity squared in the equations of motion~\eqref{eq:EffRelEQM}, we calculate the zeroth-order contribution from the radial modes 
\begin{align}
\dot{\myZOT{x}}^2 + \dot{\myZOT{y}}^2 
 &= (\myompw\myAmpR{+})^2 + (\myommw\myAmpR{-})^2 
+ 2\myompw\myommw\myAmpR{+}\myAmpR{-}\cos(\myPhiTotw{\mybeat}) 
\label{eq:xvq+yvq} \myGSAbstand,
\end{align}
where we have defined $\myPhiTotw{\mybeat}= \myPhiTotw{+} - \myPhiTotw{-}$. 
The radial contribution to the zeroth-order velocity squared oscillates at the frequency
\begin{gather}
\myombw = \myompw -\myommw 
\label{eq:myombw} \myGSAbstand .
\end{gather}

We introduce the piece of notation~$\myfComp{\cdot}{\omega}$, which retrieves the oscillatory term at the frequency~$\omega$ from the argument given in angle brackets. 
Applying the notation to \myEqua{eq:xvq+yvq}, we have
\begin{align}
\myfComp{\dot{\myZOT{x}}^2 + \dot{\myZOT{y}}^2}{\myfZero} &= 
(\myompw\myAmpR{+})^2 + (\myommw\myAmpR{-})^2 
\label{eq:xvq+yvqZero}\\
\intertext{for the constant component and}
\myfComp{\dot{\myZOT{x}}^2 + \dot{\myZOT{y}}^2}{\myombw} &= 
2\myompw\myommw\myAmpR{+}\myAmpR{-}\cos(\myPhiTw{\mybeat}) 
\label{eq:xvq+yvqBeat}
\end{align}
for the oscillatory component at the difference frequency~$\myombw$ of the radial modes. Note that the oscillatory term as well as its amplitude are recovered. 

We also introduce the short-hand notation
\begin{align}
\myZOT{x}_{\pm}
 &= \myfComp{\myZOT{x}}{\myompmw}
 = \myAmpR{\pm}\cos(\myPhiTw{\pm}) 
\label{eq:x0pm} \myGSAbstand ,\\
\myZOT{y}_{\pm}
 &= \myfComp{\myZOT{y}}{\myompmw} 
 = -\myAmpR{\pm}\sin(\myPhiTw{\pm})
 \label{eq:y0pm}
\end{align}
for the two radial eigenmotions as well as their associated velocities
\begin{align}
\dot{\myZOT{x}}_{\pm} 
 &= \myfComp{\dot{\myZOT{x}}}{\myompmw} 
 = -\myompmw\myAmpR{\pm}\sin(\myPhiTw{\pm}) 
\label{eq:xv0pm} \myGSAbstand ,\\
\dot{\myZOT{y}}_{\pm} 
&= \myfComp{\dot{\myZOT{y}}}{\myompmw} 
= -\myompmw\myAmpR{\pm}\cos(\myPhiTw{\pm}) 
\label{eq:yv0pm} \myGSAbstand .
\end{align}

Apart from the two terms with velocity squared, \myEqua{eq:EffRelEQM} also features a product of velocities and coordinates, which can be rewritten as 
\begin{gather}
\dot{x} x = \frac{1}{2}\frac{\myd }{\myd t}x^2 
\label{eq:xvxdt}
\end{gather}
with a time-derivative. 
Here we show $x$ as an example, but the result carries over to $y$ and $z$.
Inserting the zeroth-order solution~$\myZOT{x}$ from \myEqua{eq:x0} on the right hand-side produces a number of oscillatory terms and possibly a constant term. 
However, the constant term is removed by taking the time-derivative, whereas the frequencies of the oscillatory terms are unaffected by this operation. 
Consequently, there is no constant contribution: 
\begin{gather}
\myfComp{\dot{\myZOT{x}}\myZOT{x}}{\myfZero} 
 = \myfComp{\dot{\myZOT{y}}\myZOT{y}}{\myfZero} 
 = \myfComp{\dot{\myZOT{z}}\myZOT{z}}{\myfZero} 
 = 0 
 \label{eq:cooXvelZero} \myGSAbstand. 
\end{gather}
Using \myEquas{eq:x0}, \eqref{eq:y0}, \eqref{eq:xv0}, and \eqref{eq:yv0} for the zeroth-order solutions of the radial modes yields
\begin{align}
\dot{\myZOT{x}} \myZOT{x} + \dot{\myZOT{y}} \myZOT{y} 
 &= -\myAmpR{+}\myAmpR{-} 
 \left(\myompw-\myommw\right) 
 \sin(\myPhiTotw{+}-\myPhiTotw{-})
 \label{eq:xvx+yvy} 
\end{align}
for the remaining oscillatory term. 
Indeed, there is no constant contribution in the sum of the two radial terms as \myEqua{eq:cooXvelZero} predicts for the individual components alone.

\subsection{Axial mode}

The first-order relativistic axial equation of motion is given by the third component of \myEqua{eq:EffRelEQM}.
As outlined in \mySec{sec:TaM}, we will now insert the zeroth-order solutions from \myEquas{eq:x0}--\eqref{eq:z0} for the trajectory and \myEquas{eq:xv0}--\eqref{eq:zv0} for the velocities. 
Fortunately, the identities we have derived at the beginning of this section will reduce the effort of identifying the resonant terms at the axial frequency that cause a frequency-shift.

Naturally-resonant contributions at the axial frequency that result from the last term in the axial equation of motion are written as
\begin{align}
\myfComp{ 
  \vphantom{\dot{\myZOT{z}}^2}\dot{\myZOT{z}} 
	\left( 
	  \dot{\myZOT{x}}\myZOT{x} + \dot{\myZOT{y}}\myZOT{y} - 2\dot{\myZOT{z}}\myZOT{z} 
	\right)}{\myomzw}
&= \myfComp{
\vphantom{\dot{\myZOT{z}}^2} 
\dot{\myZOT{z}}}{\myomzw} 
\myfComp{\dot{\myZOT{x}}\myZOT{x} + \dot{\myZOT{y}}\myZOT{y}}{\myfZero}
 - 2\myfComp{\myZOT{z}\dot{\myZOT{z}}^2}{\myomzw} 
\label{eq:axComp1}
\end{align}
in our notation. 
With naturally resonant we mean that no specific assumptions about the relation between the frequencies are required for a term to become resonant with one eigenmode. 
According to \myEqua{eq:xvx+yvy}, the contribution from the radial modes oscillates at the difference frequency~$\myombw$, which generally is not an integer of the axial frequency. 
Therefore, mixing an oscillatory term of the radial modes with an oscillatory term at the axial frequency results in a nonresonant contribution. 
We would need a constant contribution from the radial modes for $\dot{\myZOT{z}}$ to stay resonant at the axial frequency~$\myomzw$. 
However, \myEqua{eq:cooXvelZero} informs us that such a constant term does not exist, and we are left with the purely axial term in \myEqua{eq:axComp1}. 
By decomposing the relevant trigonometric functions as
\begin{align}
\cos(\myPhiTotw{z})[\sin(\myPhiTotw{z})]^2 
&= \frac{\cos(\myPhiTotw{z})-\cos(3\myPhiTotw{z})}{4} 
 \myGSAbstand ,
\end{align}
the resonant term becomes 
\begin{gather}
\myfComp{\myZOT{z} \,\dot{\myZOT{z}}^2}{\myomzw} 
= \frac{1}{4} \myomzw^2 \myAmpZ^3 \cos(\myPhiTw{z}) 
= \frac{1}{4}(\myomzw\myAmpZ)^2\, \myZOT{z} 
\label{eq:zzvq} \myGSAbstand .
\end{gather}
In the last step, we have used the axial amplitude~$\myAmpZ$ in order to write the result as proportional to the zeroth-order solution~$\myZOT{z}$ from \myEqua{eq:z0}.
In total, we have
\begin{gather}
\myfComp{ \vphantom{\dot{\myZOT{z}}^2}\dot{\myZOT{z} } 
\left(\dot{\myZOT{x}}\myZOT{x} + \dot{\myZOT{y}}\myZOT{y} - 2\dot{\myZOT{z}}\myZOT{z}\right)}{\myomzw} 
= - 2\myfComp{\myZOT{z}\dot{\myZOT{z}}^2}{\myomzw} 
 = -\frac{1}{2}(\myomzw\myAmpZ)^2\, \myZOT{z} 
\label{eq:axRes1} \myGSAbstand . 
\end{gather}

Next, we deal with the term that contains the velocity squared in the axial equation of motion. 
Naturally-resonant terms at the axial frequency are written as
\begin{align}
\myfComp{\myZOT{z}\left(\dot{\myZOT{x}}^2+\dot{\myZOT{y}}^2+\dot{\myZOT{z}}^2\right)}{\myomzw} 
&= \myfComp{\dot{\myZOT{x}}^2+\dot{\myZOT{y}}^2}{\myfZero} 
   \myfComp{\vphantom{\dot{\myZOT{x}}^2}\myZOT{z}}{\myomzw} 
	 + \myfComp{\vphantom{\dot{\myZOT{x}}^2}\myZOT{z}\,\dot{\myZOT{z}}^2}{\myomzw}
\\
&= \left[(\myompw\myAmpR{+})^2+(\myommw\myAmpR{-})^2\right] \myZOT{z} 
+ \frac{(\myomz\myAmpZ)^2}{4} \myZOT{z} 
\label{eq:axRes2} \myGSAbstand.
\end{align}
In the second step, we have used \myEqua{eq:zzvq} for the contribution by the axial mode. 
Unlike before, there is a constant term from the radial modes (see \myEqua{eq:xvq+yvqZero}), while we have dismissed the oscillatory term at the frequency~$\myombw$ along the same lines as before.

With \myEquas{eq:axRes1} and \eqref{eq:axRes2}, the effective axial equation of motion becomes
\begin{gather}
\ddot{\myZOT{z}}  + \myomz^2\myZOT{z} 
\left[ 1 - \frac{(\myompw\myAmpR{+})^2 + (\myommw\myAmpR{-})^2 + \frac{(\myomzw\myAmpZ)^2}{4}}{2c^2}
-\frac{\frac{(\myomz\myAmpZ)^2}{2}}{2c^2} \right]=0 \myGSAbstand ,
\end{gather}
which is identical to \myEqua{eq:AxialEQM-Eff} with 
\begin{gather}
\myels_{z} = -\frac{1}{2c^2} 
\left[(\myompw\myAmpR{+})^2 + (\myommw\myAmpR{-})^2 + \frac{3}{4}(\myomzw\myAmpZ)^2\right] \myGSAbstand .
\end{gather}
Finally, the parameter~$\myels_{z}$ is related to the first-order axial frequency-shift 
\begin{gather}
\frac{\Delta \myomz}{\myomz} = -\frac{1}{4c^2}
\left[ (\myomp\myAmpR{+})^2 + (\myomm\myAmpR{-})^2 + \frac{3}{4}(\myomz\myAmpZ)^2 \right]
 \label{eq:FreqShiftAxRel}
\end{gather}
via \myEqua{eq:DeltaOmz}. 
In the last step, we have switched from the perturbed frequencies~$\myomw_{i}$ to the unperturbed frequencies~$\omega_{i}$. Nevertheless, the frequency-shift is still correct to first order in the perturbation parameter $c^{-2}$. 
Since the difference between $\myomw_{i}$ and $\omega_{i}$ is at least of first order in $c^{-2}$ just like the frequency-shift, the overall effect of the substitution is at least of second order in $c^{-2}$. 

\subsection{Radial modes}

The first-order relativistic radial equations of motion are the first two components of \myEqua{eq:EffRelEQM}.
Like for the axial mode, we will insert the zeroth-order solutions from \myEquas{eq:x0}--\eqref{eq:z0} for the trajectory and \myEquas{eq:xv0}--\eqref{eq:zv0} for the velocities. 
We will also make use of the identities derived at the beginning of this section. 

First, we examine the terms associated with $x$ and $y$. Combined with the velocity squared, naturally-resonant terms at a frequency of the radial eigenmotions are written as
\begin{gather}
\myfComp{ 
\left(\dot{\myZOT{x}}^2 + \dot{\myZOT{y}}^2 + \dot{\myZOT{z}}^2\right) 
\myZOT{x}}{\myompmw}
 = \myfComp{ \left(\dot{\myZOT{x}}^2 + \dot{\myZOT{y}}^2\right) 
\myZOT{x}}{\myompmw} 
+ \myfComp{ \vphantom{\left(\dot{\myZOT{x}}^2 + \dot{\myZOT{y}}^2\right)} 
 \dot{\myZOT{z}}^2}{\myfZero} 
\myfComp{ \vphantom{\left(\dot{\myZOT{x}}^2 + \dot{\myZOT{y}}^2\right)} 
\myZOT{x}}{\myompmw} 
\label{eq:vqx} \myGSAbstand .
\end{gather} 
Again, we have assumed that mixing the axial frequency or its multiples with a frequency of the radial modes produces a nonresonant contribution. 
However, the naturally-resonant terms in $\myZOT{x}$ are preserved by multiplying them with the constant component that results from the axial mode. 
By using 
\begin{gather}
\myfComp{\dot{\myZOT{z}}^2}{\myfZero} 
 = (\myomzw\myAmpZ)^2 \myfComp{[\sin(\myPhiTotw{z})]^2}{\myfZero} 
 = \frac{1}{2}(\myomzw\myAmpZ)^2 
\label{eq:zvsq0} \myGSAbstand ,
\end{gather}
the second term on the right-hand side of \myEqua{eq:vqx}
\begin{gather}
\myfComp{\dot{\myZOT{z}}^2}{\myfZero} 
  \myfComp{ \vphantom{\tilde{\myZOT{z}}^2} \myZOT{x}}{\myompmw} 
= \frac{1}{2}(\myomzw\myAmpZ)^2 \myZOT{x}_{\pm} 
\label{eq:zvqx}
\end{gather}
is quickly dealt with. The result carries over to the second component of the radial equations of motion with the replacements $\tilde{x}\to \tilde{y}$, and $\tilde{x}_{\pm}\to \tilde{y}_{\pm}$.

The first term on the right-hand side of \myEqua{eq:vqx} needs more scrutiny because mixing an oscillatory component at the difference frequency~$\myombw$ defined in \myEqua{eq:myombw} with the oscillatory component at the radial frequency~$\myommpw$ creates an oscillatory term at the other radial frequency~$\myompmw$. 
This is expressed as
\begin{gather}
\myfComp{ \vphantom{\myZOT{x}^2} 
\cos(\myPhiTotw{+} - \myPhiTotw{-}) \cos(\myPhiTotw{\mp})}{\myompmw} 
= \frac{1}{2}\cos(\myPhiTotw{\pm}) 
\label{eq:cosbXcosmpRes} 
\end{gather}
in our notation. 
The two relevant cases for producing resonant terms at either radial frequency are then written as
\begin{align}
& \myfComp{\left(\dot{\myZOT{x}}^2+\dot{\myZOT{y}}^2\right)\myZOT{x}}{\myompmw}
= \myfComp{ \vphantom{\left(\dot{\myZOT{x}}^2\right)} 
\dot{\myZOT{x}}^2 + \dot{\myZOT{y}}^2}{\myfZero} 
\myfComp{ \vphantom{\left(\dot{\myZOT{x}}^2\right)} 
\myZOT{x}}{\myompmw} 
+ \myfComp{ 
\myfComp{ \dot{\myZOT{x}}^2 + \dot{\myZOT{y}}^2}{\myombw}
\myZOT{x}_{\mp}}{\myompmw}\\
&= \left[(\myompw\myAmpR{+})^2 + (\myommw\myAmpR{-})^2\right] \myZOT{x}_{\pm} 
 + \myompw\myommw\myAmpR{+}\myAmpR{-}\, \myAmpR{\mp}\cos(\myPhiTotw{\pm})
\\
&= \left[ (\myompw\myAmpR{+})^2 + (\myommw\myAmpR{-})^2 
+ \myompw\myommw\myAmpR{\mp}^2 \right] \myZOT{x}_{\pm} 
\label{eq:xyvqsx} 
\end{align}
with the help of \myEquas{eq:xvq+yvqZero} and \eqref{eq:xvq+yvqBeat}. 
In the last step, we have used the amplitude~$\myAmpR{\pm}$ in order to introduce the zeroth-order solution $\myZOT{x}_{\pm} = \myAmpR{\pm} \cos(\myPhiTotw{\pm})$ from \myEqua{eq:x0pm}. 
Since
\begin{gather}
\myfComp{ \vphantom{\myZOT{x}^2} %
\cos(\myPhiTotw{+} - \myPhiTotw{-}) \sin(\myPhiTotw{\mp})}{\myompmw} %
= \frac{1}{2}\sin(\myPhiTotw{\pm}) %
\label{eq:cosbXsinmpRes} \myGSAbstand ,
\end{gather}
the result also holds when $\myZOT{x}$ is replaced by $\myZOT{y}$ as in
\begin{align}
\myfComp{ \left(\dot{\myZOT{x}}^2 + \dot{\myZOT{y}}^2\right) %
\myZOT{y}}{\myompmw} %
&= \left[ (\myompw\myAmpR{+})^2 + (\myommw\myAmpR{-})^2 %
+ \myompw\myommw\myAmpR{\mp}^2\right] \myZOT{y}_{\pm} %
\label{eq:xyvqsy} \myGSAbstand .
\end{align}

Next, we deal with the term of the kind $\myvel^2\dot{y}$ in the radial equations of motion by using the same decomposition
\begin{gather}
\myfComp{ \left(\dot{\myZOT{x}}^2 + \dot{\myZOT{y}}^2 + \dot{\myZOT{z}}^2\right) 
\dot{\myZOT{y}}}{\myompmw} %
 = \myfComp{ \left(\dot{\myZOT{x}}^2 + \dot{\myZOT{y}}^2\right) 
\dot{\myZOT{y}}}{\myompmw} %
 + \myfComp{ \vphantom{\left(\dot{\myZOT{x}}^2 + \dot{\myZOT{y}}^2\right) } %
\dot{\myZOT{z}}^2}{\myfZero} %
\myfComp{ \vphantom{\left(\dot{\myZOT{x}}^2 + \dot{\myZOT{y}}^2\right)} %
\dot{\myZOT{y}}}{\myompmw} %
\label{eq:vqyv} 
\end{gather}
as before. 
With the help of \myEqua{eq:zvsq0}, the second term on the right-hand side is expressed as
\begin{gather}
\myfComp{\dot{\myZOT{z}}^2}{\myfZero} %
\myfComp{ \vphantom{\tilde{\myZOT{z}}^2} %
\dot{\myZOT{y}}}{\myompmw} %
= \frac{1}{2}(\myomzw\myAmpZ)^2 \dot{\myZOT{y}}_{\pm} %
\label{eq:zvqyv} \myGSAbstand . 
\end{gather}
This also holds true for $\dot{\myZOT{y}}$ replaced by $\dot{\myZOT{x}}$ (and $\dot{\myZOT{y}}_{\pm}$ by $\dot{\myZOT{x}}_{\pm}$). 

For the first term in \myEqua{eq:vqyv}, we have 
\begin{align}
&\myfComp{\left(\dot{\myZOT{x}}^2+\dot{\myZOT{y}}^2\right)\dot{\myZOT{y}}}{\myompmw} %
= \myfComp{ \vphantom{\left(\dot{\myZOT{x}}^2\right)} %
\dot{\myZOT{x}}^2 + \dot{\myZOT{y}}^2}{\myfZero} %
 \myfComp{ \vphantom{\left(\dot{\myZOT{x}}^2\right)} %
\dot{\myZOT{y}}}{\myompmw} %
+ \myfComp{\myfComp{\dot{\myZOT{x}}^2 +\dot{\myZOT{y}}^2}{\myombw} %
\dot{\myZOT{y}}_{\mp}}{\myompmw}\\ %
&= \left[ (\myompw\myAmpR{+})^2 + (\myommw\myAmpR{-})^2\right] %
\dot{\myZOT{y}}_{\pm} %
 - \myompw\myommw\myAmpR{+}\myAmpR{-}\, \myommpw\myAmpR{\mp}\cos(\myPhiTotw{\pm}) %
\\ %
&= \left[ (\myompw\myAmpR{+})^2 + (\myommw\myAmpR{-})^2 \right] %
\dot{\myZOT{y}}_{\pm} %
 + (\myommpw\myAmpR{\mp})^2\, \dot{\myZOT{y}}_{\pm}\\%
&= \left[ (\myompmw\myAmpR{\pm})^2 + 2(\myommpw\myAmpR{\mp})^2 \right] %
\dot{\myZOT{y}}_{\pm} %
\label{eq:xyqsyv}
\end{align}
with the help of \myEquas{eq:xvq+yvqZero} and \eqref{eq:xvq+yvqBeat}.
In the second-to-last step, we have used a factor of $-\myompmw\myAmpR{\pm}$ in order to write $\dot{\myZOT{y}}_{\pm}\ = -\myompmw \myAmpR{\pm}\cos(\myPhiTotw{\pm})$ as defined in \myEqua{eq:yv0pm}.
Using \myEqua{eq:cosbXsinmpRes}, we find essentially the same result
\begin{gather}
\myfComp{ \left(\dot{\myZOT{x}}^2+\dot{\myZOT{y}}^2 \right) %
\dot{\myZOT{x}}}{\myompmw} %
= \left[ (\myompmw\myAmpR{\pm})^2 + 2(\myommpw\myAmpR{\mp})^2 \right]%
\dot{\myZOT{x}}_{\pm} %
\label{eq:xyqsxv} 
\end{gather}
for $\dot{\myZOT{x}}$ as for $\dot{\myZOT{y}}$.

The third term that we deal with in the radial equations of motion involves a product of velocities and coordinates. 
Picking the $x$-component as an example, naturally-resonant terms at either of the two radial frequencies are expressed by
\begin{gather}
\myfComp{ \left( \dot{\myZOT{x}}\myZOT{x} + \dot{\myZOT{y}}\myZOT{y} - 2\dot{\myZOT{z}}\myZOT{z} \right) %
\dot{\myZOT{x}}}{\myompmw} %
= \myfComp{ \left(\dot{\myZOT{x}}\myZOT{x} + \dot{\myZOT{y}}\myZOT{y} \right) %
\dot{\myZOT{x}}}{\myompmw} %
- 2 \myfComp{\dot{\myZOT{z}}\myZOT{z}}{\myfZero} %
\dot{\myZOT{x}}_{\pm} %
\myGSAbstand .
\end{gather}
As evidenced by \myEqua{eq:cooXvelZero}, there is no constant contribution in $\dot{\myZOT{z}}\myZOT{z}$, and thus the second term on the right-hand side vanishes. 
The remaining contribution is evaluated further with the help of the identity
\begin{gather}
\myfComp{ \vphantom{\myZOT{x}^2} \sin(\myPhiTotw{+} - \myPhiTotw{-}) \sin(\myPhiTotw{\mp})}{\myompmw} %
= \mp \frac{1}{2}\cos(\myPhiTotw{\pm}) %
\label{eq:sinbXsinpmRes} 
\end{gather}
and \myEqua{eq:xvx+yvy}. 
Finally, this yields
\begin{align}
&\myfComp{ \left(\dot{\myZOT{x}}\myZOT{x} + \dot{\myZOT{y}}\myZOT{y}\right) %
\dot{\myZOT{x}}}{\myompmw} %
 = \myfComp{ \left(\dot{\myZOT{x}}\myZOT{x} + \dot{\myZOT{y}}\myZOT{y}\right) %
\dot{\myZOT{x}}_{\mp}}{\myompmw} \label{eq:xvxxvStart} \\%
&= \myfComp{ \vphantom{\myZOT{x}^2} %
   [-\myAmpR{+}\myAmpR{-} (\myompw - \myommw) \sin(\myPhiTotw{+} - \myPhiTotw{-})]%
	\,[-\myommpw \myAmpR{\mp} \sin(\myPhiTotw{\mp})]}{\myompmw}\\%
&= \mp \frac{1}{2} \myAmpR{+} \myAmpR{-} \myAmpR{\mp} \,\myommpw (\myompw - \myommw) \cos(\myPhiTotw{\pm})\\ %
&= \mp \frac{1}{2} \myAmpR{\mp}^2 \,\myommpw(\myompw-\myommw) \, \myZOT{x}_{\pm} %
\label{eq:xvxxv} \myGSAbstand.
\end{align}
With the help of
\begin{gather}
\myfComp{ \vphantom{\myZOT{x}^2} %
\sin(\myPhiTotw{+} - \myPhiTotw{-}) \cos(\myPhiTotw{\mp})}{\myompmw} %
= \pm\frac{1}{2}\sin(\myPhiTotw{\pm}) %
\myGSAbstand ,
\end{gather}
the $y$-component gives
\begin{align}
&\myfComp{ \left(\dot{\myZOT{x}}\myZOT{x} + \dot{\myZOT{y}}\myZOT{y}\right) %
\dot{\myZOT{y}}}{\myompmw} 
= \myfComp{ \left(\dot{\myZOT{x}}\myZOT{x} + \dot{\myZOT{y}}\myZOT{y}\right)%
  \dot{\myZOT{y}}_{\mp}}{\myompmw} \label{eq:xvxyvStart}\\ %
&= \myfComp{ \vphantom{\myZOT{x}^2} %
[-\myAmpR{+} \myAmpR{-}(\myompw - \myommw) \sin(\myPhiTotw{+} - \myPhiTotw{-})]%
\,[-\myommpw \myAmpR{\mp} \cos(\myPhiTotw{\mp})]}{\myompmw}\\%
&= \pm \frac{1}{2} \myAmpR{+} \myAmpR{-} \myAmpR{\mp} \, \myommpw (\myompw - \myommw) \sin(\myPhiTotw{\pm})\\ %
& = \mp \frac{1}{2} \myAmpR{\mp}^2 \,\myommpw (\myompw - \myommw)\, \myZOT{y}_{\pm} %
 \label{eq:xvxyv} \myGSAbstand .
\end{align}
In both cases, we have absorbed factors of  $\myAmpR{\pm}$ and $-\myAmpR{\pm}$ in $\myZOT{x}_{\pm}$ and $\myZOT{y}_{\pm}$ from \myEquas{eq:x0pm} and~\eqref{eq:y0pm}, respectively. 
Note that the final result is proportional to $\myZOT{x}_{\pm}$ and $\myZOT{y}_{\pm}$, whereas the initial expressions came with a common factor of $\dot{\myZOT{x}}$ and $\dot{\myZOT{y}}$, respectively.
It is this change from velocities to coordinates that leads to a term%
\footnote{%
Since the terms in \myEquas{eq:xvxxvStart} and \eqref{eq:xvxyvStart} are associated with a factor of the axial frequency squared in the equations of motion~\eqref{eq:EffRelEQM}, it is only natural to write the resulting resonant terms as proportional to $\myZOT{x}_{\pm}$ and $\myZOT{y}_{\pm}$, thereby staying in line with the effective equations of motion~\eqref{eq:radEQM-Eff}, which have the free-space cyclotron-frequency~$\myomc$ associated with velocities, whereas the axial frequency~$\myomz$ appears in combination with coordinates. 
The choice of coordinates over velocities in the above case is not mandatory, however. 
\myEquas{eq:x0pm}--\eqref{eq:yv0pm} link the zeroth-order coordinates and velocities of each radial mode as $\dot{\myZOT{x}}_{\pm} = \myompmw\myZOT{y}_{\pm}$ and $\dot{\myZOT{y}}_{\pm} = -\myompmw\myZOT{x}_{\pm}$. 
Therefore, a contribution by $\myels_{\pm}$ to the equations of motion~\eqref{eq:radEQM-Eff} is the same as by $\mymfls_{\pm} = -(\myomp\myomm)/(\myompmw\myomc)\myels_{\pm}$. 
As a consistency check, both parameters yield the same first-order frequency-shift in \myEqua{eq:RadFoFs}.
} %
 that fits into the effective equations of motion~\eqref{eq:radEQM-Eff}.

Combining \myEquas{eq:zvqx}, \eqref{eq:xyvqsx}, \eqref{eq:xyvqsy}, \eqref{eq:zvqyv}, \eqref{eq:xyqsyv}, \eqref{eq:xyqsxv}, \eqref{eq:xvxxv}, and \eqref{eq:xvxyv}, the effective radial equations of motion become
\begin{gather}
\begin{split}
\begin{pmatrix}\ddot{\myZOT{x}}_{\pm}\\ \ddot{\myZOT{y}}_{\pm}\end{pmatrix} 
& = \left[ 1 - \frac{(\myompmw\myAmpR{\pm})^2 + 2(\myommpw\myAmpR{\mp})^2 + \frac{1}{2}(\myomzw\myAmpZ)^2}{2c^2} \right] \myomc %
\begin{pmatrix} \dot{\myZOT{y}}_{\pm}\\ -\dot{\myZOT{x}}_{\pm}\end{pmatrix} \\
 & \quad {}+ %
\left[ 1- \frac{(\myompw\myAmpR{+})^2 + (\myommw\myAmpR{-})^2 + \myompw \myommw \myAmpR{\mp}^2 + \frac{1}{2}(\myomzw\myAmpZ)^2}{2c^2}\right. %
\\ %
& \quad\quad\quad%
- \left.  \frac{\mp\myAmpR{\mp}^2\,\myommpw(\myompw-\myommw)}{2c^2}\right] %
 \frac{\myomz^2}{2}
\begin{pmatrix}\myZOT{x}_{\pm}\\ \myZOT{y}_{\pm} \end{pmatrix}
\label{eq:EffRadEQMStep1} \myGSAbstand .
\end{split}
\end{gather}
%
After simplifying the above equation using 
\begin{gather}
\mp \myommpw(\myompw-\myommw) = \myommpw^2-\myompw\myommw %
\myGSAbstand ,
\end{gather}
the parameters in \myEqua{eq:radEQM-Eff} are identified as 
\begin{gather}
\mymfls_{\pm} %
= \myels_{\pm} %
= -\frac{1}{2c^2} %
\left[ (\myompmw\myAmpR{\pm})^2 + 2(\myommpw\myAmpR{\mp})^2 + \frac{1}{2}(\myomzw\myAmpZ)^2 \right] %
\myGSAbstand.
\end{gather}
These are related to the first-order frequency-shift via Equation~\eqref{eq:RadFoFs}, whose numerator
\begin{gather}
\pm \myompm\myomc \mp \myomp\myomm %
= \pm \myompm(\myomp+\myomm) \mp \myomp\myomm  %
= \pm \myompm^2 %
\label{eq:sum01}
\end{gather}
we simply add, since the two parameters~$\mymfls_{\pm}$ and $\myels_{\pm}$ are equal. In the process, we have used the sideband identity from \myEqua{eq:sidebandSum}. 
Thus, we obtain
\begin{gather}
\Delta \myompm %
= \pm \frac{\myompm^2}{\myomp-\myomm}\mymfls_{\pm} %
= \pm \frac{\myompm^2}{\myomp-\myomm}\myels_{\pm} %
\label{eq:FreqShiftRadPara} 
\end{gather}
for the general relation and 
\begin{gather}
\frac{\Delta \myompm}{\myompm} %
= \mp\frac{\myompm}{\myomp-\myomm} %
\frac{(\myompm\myAmpR{\pm})^2 + 2(\myommp\myAmpR{\mp})^2+ \frac{1}{2}(\myomz\myAmpZ)^2}{2c^2} %
\label{eq:FreqShiftRadRel} 
\end{gather}
for the first-order frequency-shift. 
Like for the axial mode, we have switched back to the unperturbed frequencies~$\omega_{i}$ here, which does not affect the outcome to first order in the perturbation parameter $c^{-2}$.

\subsection{Comparison}

For easier comparison with other results, we will express the frequency-shifts as a function of the energies
\begin{align}
\myEn{+} &= %
\frac{1}{2} m \myomp(\myomp-\myomm)\myAmpR{+}^2 %
\approx \frac{1}{2}m\myomp^2\myAmpR{+}^2 %
\label{eq:Ecyc} \myGSAbstand , \\
\myEn{-} &= %
-\frac{1}{2}m \myomm(\myomp-\myomm)\myAmpR{-}^2 %
\approx -\frac{1}{4}m\myomz^2\myAmpR{-}^2 
\label{eq:Emag} \myGSAbstand ,\\
\myEn{z} &= \frac{1}{2}m\myomz^2\myAmpZ^2
\label{eq:Eaxial}
\end{align}
associated with the three eigenmodes of a charged particle in an ideal Penning trap. For the approximate expressions, we have assumed the typical hierarchy of $\lvert\myomp\rvert \gg \lvert\myomm\rvert$, which results in $\myomp-\myomm \approx \myomp$.
The approximation for the energy of the magnetron mode also uses \myEqua{eq:radProd}, $\myomz^2=2\myomp\myomm$, which is equally important for the comparison of our result with the literature. 

Substituting the full expressions for the energies from \myEquas{eq:Ecyc}--\eqref{eq:Eaxial} for the amplitudes in \myEquas{eq:FreqShiftAxRel} and \eqref{eq:FreqShiftRadRel} yields 
\begin{align}
\frac{\Delta \myomp}{\myomp} &= %
\frac{-1}{mc^2} %
\left[ \frac{\myomp^2\, \myEn{+}}{(\myomp-\myomm)^2} %
+ \frac{\myomp\, \myEn{z}}{2(\myomp-\myomm)} %
- \frac{\myomz^2\, \myEn{-}}{(\myomp-\myomm)^2} %
\right]  %
\myGSAbstand,\\%
\frac{\Delta \myomz}{\myomz} &= %
\frac{-1}{mc^2} %
\left[ \frac{\myomp\, \myEn{+}}{2(\myomp-\myomm)} %
+ \frac{3}{8}\myEn{z} %
- \frac{\myomz^2\, \myEn{-}}{4\myomp(\myomp-\myomm)} %
\right] %
\myGSAbstand, \\ %
\frac{\Delta \myomm}{\myomm} &= %
\frac{1}{mc^2} %
\left[ \frac{\myomz^2\, \myEn{+}}{(\myomp-\myomm)^2} %
 + \frac{\myomm\, \myEn{z}}{2(\myomp-\myomm)} %
 - \frac{\myomm^2\, \myEn{-}}{(\myomp-\myomm)^2} %
\right] %
\end{align}
for the frequency-shifts. 
Neglecting $\myomm$ against $\myomp$, the result agrees with the classical limit given in~\cite{brown1986}. 
The fully quantum-mechanical result from \cite{brown1986} was expressed as a function of the energies~$\myEn{i}$ in~\cite{farnham1995} without any particular assumption about the frequencies.  
Ignoring spin and the zero-point shift that results from the nonzero energy in the quantum-mechanical ground-state, our classical result is in agreement, too.

\section{Estimates based on relativistic mass-increase} \label{sec:EffMass}

In this section, we investigate how well the frequency-shifts can be understood in terms of relativistic mass-increase. 
This simple model is widely used to estimate and explain the relativistic frequency-shifts~\cite{brown1986}, although \myEqua{eq:RelEQM} shows that the relativistic corrections to the equations of motion are more complex than just the Lorentz factor~$\gamma$. 
Nevertheless, the simple model is surprisingly accurate. 

For small velocities, the relativistic mass of the particle is approximated by
\begin{gather}
\gamma m %
= \frac{m}{\sqrt{1-\myvel^2/c^2}} %
\approx m\left(1+ \frac{\myvel^2}{2c^2} + \cdots \right) %
\end{gather}
with the mass-increase
\begin{gather}
\frac{\Delta m}{m} = \frac{\myvel^2}{2c^2} %
\label{eq:MassIncVel} \myGSAbstand .
\end{gather}
By using the dependence of the frequency on mass, this mass-increase~$\Delta m$ is translated into a frequency-shift. 

The mass-dependence of the free-space cyclotron-frequency from \myEqua{eq:omc} is
\begin{gather}
\frac{\myd \myomc}{\myd m} = - \frac{\myomc}{m} %
\myGSAbstand ,
\end{gather}
which results in the prediction
\begin{gather}
\frac{\Delta \myomc}{\myomc} %
= - \frac{\Delta m}{m} %
= -\frac{\myvel^2}{2c^2} %
= -\frac{(\myomc \myAmpR{\mathrm{c}})^2}{2c^2}
\end{gather}
for the frequency-shift. 
In the last step, we have used the velocity squared of a purely circular motion at the free-space cyclotron-frequency~$\myomc$ with the cyclotron radius~$\myAmpR{\mathrm{c}}$, thereby effectively ignoring the axial motion. 
With no electric field, axial energy would lead to a loss of the particle anyway. 
Since experiments are often performed in the regime of $\lvert\myomc\rvert \gtrsim \lvert \myomp\rvert \gg \myomz \gg \lvert\myomm\rvert$, the prediction for~$\myomc$ is assumed to be valid for the modified cyclotron-frequency~$\myomp$, too. 

We can strengthen the argument by calculating the mass-dependence 
\begin{gather}
\frac{\myd\myompm}{\myd m} = \mp\frac{1}{m}\frac{\myompm^2}{\myomp-\myomm}
\end{gather}
of the two radial frequencies from \myEqua{eq:ompm0rad}. 
With \myEqua{eq:MassIncVel} our estimate for the frequency-shift then becomes
\begin{gather}
\frac{\Delta \myompm}{\myompm} %
= \mp\frac{\myompm}{\myomp-\myomm} \frac{\Delta m}{m} %
= \mp\frac{\myompm}{\myomp-\myomm} \frac{\myvel^2}{2c^2} %
\label{eq:ompmShiftEstimate} \myGSAbstand . 
\end{gather}
From \myEquas{eq:xvq+yvqZero} and \eqref{eq:zvsq0}, we take the zeroth-order estimate for the mean of the velocity squared as
\begin{gather}
\myfComp{\myZOT{\myvel}^2}{\myfZero} %
= (\myompw\myAmpR{+})^2 + (\myommw\myAmpR{-})^2 %
+ \frac{1}{2}(\myomzw\myAmpZ)^2 %
\label{eq:vqZero} \myGSAbstand .
\end{gather}
To first order in the estimated frequency-shift, we can replace the perturbed frequencies~$\myomw_{i}$ with the unperturbed ones when plugging the result into \myEqua{eq:ompmShiftEstimate}. 
Altogether, the estimate is almost equal to the actual relativistic shifts given in \myEqua{eq:FreqShiftRadRel}, apart from missing a factor of~2 in the dependence of the radial frequency~$\myompmw$ on the amplitude squared of the other radial motion. 

The mass-dependence
\begin{gather}
\frac{\myd \myomz}{\myd m}  = -\frac{\myomz}{2m} 
\end{gather}
of the axial frequency defined in \myEqua{eq:omz} results in the estimate
\begin{gather}
\frac{\Delta \myomz}{\myomz} %
= -\frac{\Delta m}{2m} %
= -\frac{\myvel^2}{4c^2} 
\end{gather}
for the relativistic frequency-shift. 
By inserting \myEqua{eq:vqZero} with the substitution~$\myomw_{i}\to \omega_{i}$, the actual dependence on the amplitudes of the radial modes is correctly reproduced. 
However, the exact result given in \myEqua{eq:FreqShiftAxRel} depends more strongly on the axial amplitude squared by a factor of $3/2$. 

The pitfalls of the simple model based on relativistic mass-increase are summarized in \myTab{tab:tab01}. 
Given the aforementioned hierarchy of frequencies, the modified cyclotron-motion is subject to the largest relativistic frequency-shifts, and it is also the strongest contributor to them, assuming similar motional amplitudes. 
Fortunately, the most relevant relativistic shift in many Penning-trap experiments---the dependence of the reduced cyclotron-frequency~$\myompw$ on $\myAmpR{+}$---is predicted correctly by the simple model, whose flaws happen to afflict less important contributions. 

\begin{table}
\caption{%
Comparison of the simple estimate based on relativistic mass-increase with the exact first-order calculation for the dependence of the relativistic frequency-shift~$\Delta \omega_{i}$ on the amplitudes of the three eigenmotions. %
The checkmark~($\checkmark$) indicates agreement; %
the cross~($\times$) indicates a discrepancy. %
}
\centering
\begin{tabular}{l|ccc}
 & $\myAmpR{+}$ & $\myAmpZ$ & $\myAmpR{-}$\\ \hline
$\Delta \myomp$ & $\checkmark$ & $\checkmark$ & $\times$\\
$\Delta \myomz$ & $\checkmark$ & $\times$ & $\checkmark$ \\
$\Delta \myomm$ & $\times$ & $\checkmark$ & $\checkmark$\\
\end{tabular}
\label{tab:tab01}
\end{table}

\section{Conclusion}
Using an adequate approximation of the relativistic equation of motion in the ideal Penning trap, we have calculated the first-order relativistic frequency-shifts caused by the motional degrees of freedom. 
Quantization aside, the result agrees with previous treatments via operator formalisms. 
To say the least, a simple model of relativistic mass-increase often produces more than the right order of magnitude for the expected frequency-shift. 
There is perfect agreement with the first-order result in six out of nine dependencies.

\section*{Acknowledgments}

This work was funded by the Max-Planck-Gesellschaft and the ERC Grant Precision Measurements of Fundamental Constants (MEFUCO). 
T.\,E.\ was supported by a fellowship of the Alexander von Humboldt foundation. S.\,S.\ acknowledges support by the Heidelberg Graduate School of Fundamental Physics (HGSFP). 
J.\,K.\ acknowledges support by the HGSFP and by the International Max Planck Research School for Precision Tests of Fundamental Symmetries (IMPRS-PTFS). 



\begin{thebibliography}{19}
\expandafter\ifx\csname natexlab\endcsname\relax\def\natexlab#1{#1}\fi
\providecommand{\url}[1]{\texttt{#1}}
\providecommand{\href}[2]{#2}
\providecommand{\path}[1]{#1}
\providecommand{\DOIprefix}{doi:}
\providecommand{\ArXivprefix}{arXiv:}
\providecommand{\URLprefix}{URL: }
\providecommand{\Pubmedprefix}{pmid:}
\providecommand{\doi}[1]{\href{http://dx.doi.org/#1}{\path{#1}}}
\providecommand{\Pubmed}[1]{\href{pmid:#1}{\path{#1}}}
\providecommand{\bibinfo}[2]{#2}
\ifx\xfnm\relax \def\xfnm[#1]{\unskip,\space#1}\fi
\bibitem[{Blaum et~al.(2010)Blaum, Novikov, and Werth}]{blaum2010}
\bibinfo{author}{K.~Blaum}, \bibinfo{author}{{\relax Yu}.~N. Novikov},
  \bibinfo{author}{G.~Werth},
\newblock \bibinfo{title}{{P}enning traps as a versatile tool for precise
  experiments in fundamental physics},
\newblock \bibinfo{journal}{Contemporary Physics} \bibinfo{volume}{51}
  (\bibinfo{year}{2010}) \bibinfo{pages}{149--175}. \URLprefix
  \url{http://www.tandfonline.com/doi/abs/10.1080/00107510903387652}.
  \DOIprefix\doi{10.1080/00107510903387652}.
\bibitem[{Gabrielse et~al.(1995)Gabrielse, Phillips, Quint, Kalinowsky,
  Rouleau, and Jhe}]{gabrielse1995}
\bibinfo{author}{G.~Gabrielse}, \bibinfo{author}{D.~Phillips},
  \bibinfo{author}{W.~Quint}, \bibinfo{author}{H.~Kalinowsky},
  \bibinfo{author}{G.~Rouleau}, \bibinfo{author}{W.~Jhe},
\newblock \bibinfo{title}{Special relativity and the single antiproton:
  Fortyfold improved comparison of $\overline{p}$ and $p$ charge-to-mass
  ratios},
\newblock \bibinfo{journal}{Physical Review Letters} \bibinfo{volume}{74}
  (\bibinfo{year}{1995}) \bibinfo{pages}{3544--3547}. \URLprefix
  \url{http://link.aps.org/doi/10.1103/PhysRevLett.74.3544}.
  \DOIprefix\doi{10.1103/PhysRevLett.74.3544}.
\bibitem[{Bergstr{\"o}m et~al.(2002)Bergstr{\"o}m, Carlberg, Fritioff,
  Douysset, Sch{\"o}nfelder, and Schuch}]{bergstroem2002}
\bibinfo{author}{I.~Bergstr{\"o}m}, \bibinfo{author}{C.~Carlberg},
  \bibinfo{author}{T.~Fritioff}, \bibinfo{author}{G.~Douysset},
  \bibinfo{author}{J.~Sch{\"o}nfelder}, \bibinfo{author}{R.~Schuch},
\newblock \bibinfo{title}{{SMILETRAP}---{A} {P}enning trap facility for
  precision mass measurements using highly charged ions},
\newblock \bibinfo{journal}{Nuclear Instruments and Methods in Physics Research
  Section A: Accelerators, Spectrometers, Detectors and Associated Equipment}
  \bibinfo{volume}{487} (\bibinfo{year}{2002}) \bibinfo{pages}{618--651}.
  \URLprefix
  \url{http://www.sciencedirect.com/science/article/pii/S0168900201021787}.
  \DOIprefix\doi{10.1016/S0168-9002(01)02178-7}.
\bibitem[{Brodeur et~al.(2009)Brodeur, Brunner, Champagne, Ettenauer, Smith,
  Lapierre, Ringle, Ryjkov, Audi, Delheij, Lunney, and Dilling}]{brodeur2009}
\bibinfo{author}{M.~Brodeur}, \bibinfo{author}{T.~Brunner},
  \bibinfo{author}{C.~Champagne}, \bibinfo{author}{S.~Ettenauer},
  \bibinfo{author}{M.~Smith}, \bibinfo{author}{A.~Lapierre},
  \bibinfo{author}{R.~Ringle}, \bibinfo{author}{V.~L. Ryjkov},
  \bibinfo{author}{G.~Audi}, \bibinfo{author}{P.~Delheij},
  \bibinfo{author}{D.~Lunney}, \bibinfo{author}{J.~Dilling},
\newblock \bibinfo{title}{New mass measurement of $^6${L}i and ppb-level
  systematic studies of the {P}enning trap mass spectrometer {TITAN}},
\newblock \bibinfo{journal}{Physical Review C} \bibinfo{volume}{80}
  (\bibinfo{year}{2009}) \bibinfo{pages}{044318}. \URLprefix
  \url{http://link.aps.org/doi/10.1103/PhysRevC.80.044318}.
  \DOIprefix\doi{10.1103/PhysRevC.80.044318}.
\bibitem[{Sturm et~al.(2011)Sturm, Wagner, Schabinger, and Blaum}]{sturm2011}
\bibinfo{author}{S.~Sturm}, \bibinfo{author}{A.~Wagner},
  \bibinfo{author}{B.~Schabinger}, \bibinfo{author}{K.~Blaum},
\newblock \bibinfo{title}{Phase-sensitive cyclotron frequency measurements at
  ultralow energies},
\newblock \bibinfo{journal}{Physical Review Letters} \bibinfo{volume}{107}
  (\bibinfo{year}{2011}) \bibinfo{pages}{143003}. \URLprefix
  \url{http://link.aps.org/doi/10.1103/PhysRevLett.107.143003}.
  \DOIprefix\doi{10.1103/PhysRevLett.107.143003}.
\bibitem[{Redshaw et~al.(2006)Redshaw, McDaniel, Shi, and Myers}]{redshaw2006}
\bibinfo{author}{M.~Redshaw}, \bibinfo{author}{J.~McDaniel},
  \bibinfo{author}{W.~Shi}, \bibinfo{author}{E.~G. Myers},
\newblock \bibinfo{title}{Mass ratio of two ions in a {P}enning trap by
  alternating between the trap center and a large cyclotron orbit},
\newblock \bibinfo{journal}{International Journal of Mass Spectrometry}
  \bibinfo{volume}{251} (\bibinfo{year}{2006}) \bibinfo{pages}{125--130}.
  \URLprefix
  \url{http://www.sciencedirect.com/science/article/pii/S1387380606000492}.
  \DOIprefix\doi{10.1016/j.ijms.2006.01.015}.
\bibitem[{Myers(2013)}]{myers2013}
\bibinfo{author}{E.~G. Myers},
\newblock \bibinfo{title}{The most precise atomic mass measurements in
  {P}enning traps},
\newblock \bibinfo{journal}{International Journal of Mass Spectrometry}
  \bibinfo{volume}{349--350} (\bibinfo{year}{2013}) \bibinfo{pages}{107--122}.
  \URLprefix
  \url{http://www.sciencedirect.com/science/article/pii/S1387380613001097}.
  \DOIprefix\doi{10.1016/j.ijms.2013.03.018}.
\bibitem[{Van~Dyck et~al.(1986)Van~Dyck, Schwinberg, and Dehmelt}]{vanDyck1986}
\bibinfo{author}{R.~S. Van~Dyck, Jr.}, \bibinfo{author}{P.~B. Schwinberg},
  \bibinfo{author}{H.~G. Dehmelt},
\newblock \bibinfo{title}{Electron magnetic moment from geonium spectra: Early
  experiments and background concepts},
\newblock \bibinfo{journal}{Physical Review D} \bibinfo{volume}{34}
  (\bibinfo{year}{1986}) \bibinfo{pages}{722--736}. \URLprefix
  \url{http://link.aps.org/doi/10.1103/PhysRevD.34.722}.
  \DOIprefix\doi{10.1103/PhysRevD.34.722}.
\bibitem[{Gr{\"a}ff et~al.(1969)Gr{\"a}ff, Klempt, and Werth}]{graeff1969}
\bibinfo{author}{G.~Gr{\"a}ff}, \bibinfo{author}{E.~Klempt},
  \bibinfo{author}{G.~Werth},
\newblock \bibinfo{title}{Method for measuring the anomalous magnetic moment of
  free electrons},
\newblock \bibinfo{journal}{Zeitschrift f{\"u}r Physik} \bibinfo{volume}{222}
  (\bibinfo{year}{1969}) \bibinfo{pages}{201--207}. \URLprefix
  \url{http://dx.doi.org/10.1007/BF01392119}.
  \DOIprefix\doi{10.1007/BF01392119}.
\bibitem[{Ringhofer(1974)}]{ringhofer1974}
\bibinfo{author}{K.~Ringhofer},
\newblock \bibinfo{title}{``{C}lassical'' treatment of $(g-2)$-resonance
  experiments},
\newblock \bibinfo{journal}{Acta Physica Austriaca} \bibinfo{volume}{39}
  (\bibinfo{year}{1974}) \bibinfo{pages}{193--197}.
\bibitem[{Brown and Gabrielse(1986)}]{brown1986}
\bibinfo{author}{L.~S. Brown}, \bibinfo{author}{G.~Gabrielse},
\newblock \bibinfo{title}{Geonium theory: Physics of a single electron or ion
  in a {P}enning trap},
\newblock \bibinfo{journal}{Reviews of Modern Physics} \bibinfo{volume}{58}
  (\bibinfo{year}{1986}) \bibinfo{pages}{233--311}. \URLprefix
  \url{http://link.aps.org/doi/10.1103/RevModPhys.58.233}.
  \DOIprefix\doi{10.1103/RevModPhys.58.233}.
\bibitem[{Major et~al.(2005)Major, Gheorghe, and Werth}]{cptI2005}
\bibinfo{author}{F.~G. Major}, \bibinfo{author}{V.~N. Gheorghe},
  \bibinfo{author}{G.~Werth}, \bibinfo{title}{Charged Particle Traps},
  \bibinfo{publisher}{Springer}, \bibinfo{address}{Berlin Heidelberg},
  \bibinfo{year}{2005}. \URLprefix
  \url{http://link.springer.com/book/10.1007/b137836}.
  \DOIprefix\doi{10.1007/b137836}.
\bibitem[{Kaplan(1982)}]{kaplan1982}
\bibinfo{author}{A.~E. Kaplan},
\newblock \bibinfo{title}{Hysteresis in cyclotron resonance based on weak
  relativistic-mass effects of the electron},
\newblock \bibinfo{journal}{Physical Review Letters} \bibinfo{volume}{48}
  (\bibinfo{year}{1982}) \bibinfo{pages}{138--141}. \URLprefix
  \url{http://link.aps.org/doi/10.1103/PhysRevLett.48.138}.
  \DOIprefix\doi{10.1103/PhysRevLett.48.138}.
\bibitem[{Gabrielse et~al.(1985)Gabrielse, Dehmelt, and Kells}]{gabrielse1985}
\bibinfo{author}{G.~Gabrielse}, \bibinfo{author}{H.~Dehmelt},
  \bibinfo{author}{W.~Kells},
\newblock \bibinfo{title}{Observation of a relativistic, bistable hysteresis in
  the cyclotron motion of a single electron},
\newblock \bibinfo{journal}{Physical Review Letters} \bibinfo{volume}{54}
  (\bibinfo{year}{1985}) \bibinfo{pages}{537--539}. \URLprefix
  \url{http://link.aps.org/doi/10.1103/PhysRevLett.54.537}.
  \DOIprefix\doi{10.1103/PhysRevLett.54.537}.
\bibitem[{Guan et~al.(1993)Guan, Gorshkov, Alber, and Marshall}]{guan1993}
\bibinfo{author}{S.~Guan}, \bibinfo{author}{M.~V. Gorshkov},
  \bibinfo{author}{G.~M. Alber}, \bibinfo{author}{A.~G. Marshall},
\newblock \bibinfo{title}{Resonant excitation of relativistic-ion cyclotron
  orbital motion},
\newblock \bibinfo{journal}{Physical Review A} \bibinfo{volume}{47}
  (\bibinfo{year}{1993}) \bibinfo{pages}{2730--2737}. \URLprefix
  \url{http://link.aps.org/doi/10.1103/PhysRevA.47.2730}.
  \DOIprefix\doi{10.1103/PhysRevA.47.2730}.
\bibitem[{Ketter et~al.(2014)Ketter, Eronen, H{\"o}cker, Streubel, and
  Blaum}]{ketter2013ijms}
\bibinfo{author}{J.~Ketter}, \bibinfo{author}{T.~Eronen},
  \bibinfo{author}{M.~H{\"o}cker}, \bibinfo{author}{S.~Streubel},
  \bibinfo{author}{K.~Blaum},
\newblock \bibinfo{title}{First-order perturbative calculation of the
  frequency-shifts caused by static cylindrically-symmetric electric and
  magnetic imperfections of a {P}enning trap},
\newblock \bibinfo{journal}{International Journal of Mass Spectrometry}
  \bibinfo{volume}{358} (\bibinfo{year}{2014}) \bibinfo{pages}{1--16}.
  \URLprefix
  \url{http://www.sciencedirect.com/science/article/pii/S1387380613003722}.
  \DOIprefix\doi{10.1016/j.ijms.2013.10.005}.
\bibitem[{Gr{\"a}ff and Klempt(1967)}]{graeff1967}
\bibinfo{author}{G.~Gr{\"a}ff}, \bibinfo{author}{E.~Klempt},
\newblock \bibinfo{title}{{M}essung der {Z}yklotronfrequenz im
  {V}ierpolk{\"a}fig},
\newblock \bibinfo{journal}{Zeitschrift f{\"u}r Naturforschung}
  \bibinfo{volume}{22a} (\bibinfo{year}{1967}) \bibinfo{pages}{1960--1962}.
\bibitem[{Sokolov and Pavlenko(1967)}]{sokolov1967}
\bibinfo{author}{A.~A. Sokolov}, \bibinfo{author}{{\relax Yu}.~G. Pavlenko},
\newblock \bibinfo{title}{Induced and spontaneous emission in crossed fields},
\newblock \bibinfo{journal}{Optics and Spectroscopy} \bibinfo{volume}{12}
  (\bibinfo{year}{1967}) \bibinfo{pages}{3--8}.
\bibitem[{Farnham(1995)}]{farnham1995}
\bibinfo{author}{D.~L. Farnham}, \bibinfo{title}{A Determination of the
  Proton/Electron Mass Ratio and the Electron's Atomic Mass via {P}enning Trap
  Mass Spectroscopy}, \bibinfo{type}{{Thesis (Ph. D.)}}, University of
  Washington, Seattle, \bibinfo{year}{1995}.

\end{thebibliography}
\end{document}